УДК 551.466.3

# ДВУХСОЛИТОННОЕ ВЗАИМОДЕЙСТВИЕ В РАМКАХ МОДИФИЦИРОВАННОГО УРАВНЕНИЯ КОРТЕВЕГА - ДЕ ВРИЗА

Е.Н. Пелиновский, Е.Г. Шургалина

Институт прикладной физики РАН, 603950, г. Нижний Новгород, ул. Ульянова, 46

Изучается взаимодействие двух солитонов разной полярности в рамках модифицированного уравнения Кортевега - де Вриза (мКдВ). Рассмотрено три типа взаимодействия солитонов: обменный (exchange) и обгонный (overtaking) для положительных солитонов, и поглощательно-излучательный (absorb-emit) для разнополярных солитонов. Подробно исследованы особенности взаимодействия солитонов. Поскольку взаимодействие солитонов является элементарным актом солитонной турбулентности, то изучены моменты волнового поля вплоть до четвёртого, обычно рассматриваемые в теории турбулентности. Показано, что при взаимодействии солитонов с одинаковой полярностью третий и четвертый моменты волнового поля, определяющие коэффициенты асимметрии и эксцесса в теории турбулентности, уменьшаются, в то время, как для солитонов разной полярности эти моменты увеличиваются. Полученные результаты сопоставлены с оценками для двухсолитонного взаимодействия в рамках уравнения Кортевега - де Вриза (КдВ).

**ВВЕДЕНИЕ**

Уравнение Кортевега – де Вриза (КдВ) является в настоящее время эталонным уравнением нелинейной теории волн, описывающим слабонелинейные и слабодисперсионные волны в воде и атмосфере, плазме и астрофизических системах [1]. Его солитонные и многосолитонные решения на бесконечном интервале, так же как их взаимодействия между собой и с внешними полями, очень хорошо изучены. В тоже время статистическая динамика ансамбля солитонов представляет собой еще не решенную задачу, хотя здесь получено уже довольно много численных результатов [2-6].

Модифицированное уравнение Кортевега-де Вриза (мКдВ) применяется для описания волн в изотропных средах (размерно-квантованные пленки, акустические волны в плазме, внутренние волны в симметрично стратифицированной жидкости) [1, 7-9]. С математической точки зрения оно также хорошо изучено, как и КдВ уравнение, являясь полностью интегрируемым [10]. Между тем, даже простое взаимодействие двух солитонов практически не было исследовано до самого последнего времени [11, 12], не говоря уж о статистическом описании солитонов в рамках данного уравнения. Новым моментом здесь по сравнению с КдВ-уравнением является существование солитонов обеих полярностей, взаимодействие которых имеет свои особенности.

Как известно, в теории турбулентности статистические свойства поля характеризуются первыми четырьмя моментами (среднее, дисперсия, асимметрия, эксцесс), которые легко вычисляются в измерениях [13-16]. Особенно важен четвертый момент, который определяет роль больших выбросов, и с точки зрения волновой турбулентности свидетельствует о вероятности появления волн-убийц [17]. В классической турбулентности вклад в моменты волнового поля дают взаимодействия двух, трех, четырех (и более) частиц. В теории волновой турбулентности зачастую ограничиваются волновыми взаимодействиями трех и четырех частиц [13, 18]. Специфика солитонной турбулентности в рамках уравнения Кортевега-де Вриза, как это было показано Захаровым в 1971 году [19], а затем подтверждено в работах [20-22], заключается в том, что солитоны взаимодействуют попарно. В работах [23-24] исследовано взаимодействие двух КдВ - солитонов как элементарного акта солитонной турбулентности. Подобные взаимодействия ведут к уменьшению третьего и четвёртого моментов волнового поля, в то время как первый и второй моменты, являющиеся инвариантами КдВ уравнения, остаются постоянными. Это ведёт к уменьшению асимметрии и эксцесса мультисолитонного поля.

Очевидно, что парные взаимодействия так же вносят основной вклад и в динамику мультисолитонных полей в рамках модифицированного уравнения Кортевега- де Вриза в силу его полной интегрируемости. Именно поэтому в данной работе мы сосредоточились на изучении вклада двухсолитонного взаимодействия в моменты волнового поля.

# 1 ВЗАИМОДЕЙСТВИЕ СОЛИТОНОВ

Мы будем здесь использовать каноническую форму уравнения мКдВ:

$$\frac{\partial u}{\partial t} + 6u^2 \frac{\partial u}{\partial x} + \frac{\partial^3 u}{\partial x^3} = 0, \tag{1}$$

Его точным решением является солитон:

$$u(x,t) = sA\,\text{sech}[A(x - ct - x_0)], \qquad c = A^2 \tag{2}$$

где $A$ - амплитуда солитона, $s = \pm 1$ определяет полярность солитона, $c$ - скорость солитона и $x_0$ - фаза (начальное положение солитона).

Двухсолитонное решение имеет более сложную структуру [12]:

$$u(x,t) = 2\gamma \frac{s_1 A_1 \cosh(A_2(x - A_2 t)) + s_2 A_2 \cosh(A_1(x - A_1 t))}{s_1 s_2 (\gamma^2 - 1) + \gamma^2 \cosh(A_1(x - A_1 t) - A_2(x - A_2 t)) + \cosh(A_1(x - A_1 t) + A_2(x - A_2 t))}, \quad (3)$$

$$\gamma = \frac{A_1 + A_2}{A_1 - A_2} > 1.$$

При удалении солитонов друг от друга решение (3) представляется суммой двух невзаимодействующих солитонов:

$$u(x,t) = u_1(x,t) + u_2(x,t), \quad (4)$$

где $u_{1,2}$ есть односолитонное решение (2) с амплитудами $A_{1,2}$.

Наиболее сильное взаимодействие солитонов происходит в момент их максимального сближения, чему соответствует время *t = 0*. Форма результирующего импульса легко находится из (3)

$$u(x,0) = 2\gamma \frac{A_1 \cosh(A_2 x) + s_2 A_2 \cosh(A_1 x)}{s_2(\gamma^2 - 1) + \gamma^2 \cosh[(A_1 - A_2)x] + \cosh[A_1 + A_2)x]}. \quad (5)$$

Определим форму импульса в момент взаимодействия солитонов. Как известно, при взаимодействии КдВ солитонов результирующий импульс будет одногорбый, если амплитуды солитонов сильно отличаются друг от друга, и двугорбый, если амплитуды близки [24, 25]. Аналогичные результаты можно получить для мКдВ солитонов, анализируя вторую производную функции *u(x,0)* в точке *x=0*:

$$u_{xx}(x,0)\big|_{x=0} = (A_1 - A_2)\left(A_1 A_2 - (A_1 - s_2 A_2)^2\right) \quad (6)$$

Сразу видно, что при взаимодействии солитонов разной полярности (*s₂ = - 1*) эта величина всегда отрицательна и, по крайней мере, центральная часть результирующего импульса одногорбая.

В случае же солитонов одной полярности, знак второй производной (6) меняется при

$$\frac{A_2}{A_1} = \frac{3-\sqrt{5}}{2} = 0.382, \qquad (7)$$

так что при меньшей амплитуде второго солитона результирующий импульс является одногорбым (обгонное взаимодействие, а при больших – двугорбым (обменное взаимодействие).

Итак, в мКдВ уравнении существуют три типа взаимодействия солитнов. Для разнополярных солитонов, когда быстрый солитон поглощает медленный, а затем, после прохождения быстрого через него, принимает свою первичную форму. Этот тип взаимодействия назван в [12] как absorb-emit, для положительных солитонов реализуются два типа взаимодействий: обгонный ($A_2 < 0.38\,A_1$) и обменный ($A_2 > 0.38\,A_1$).

Формы импульсов в момент взаимодействия солитонов для всех трёх случаев представлены на рисунке 1:

а) б) в)

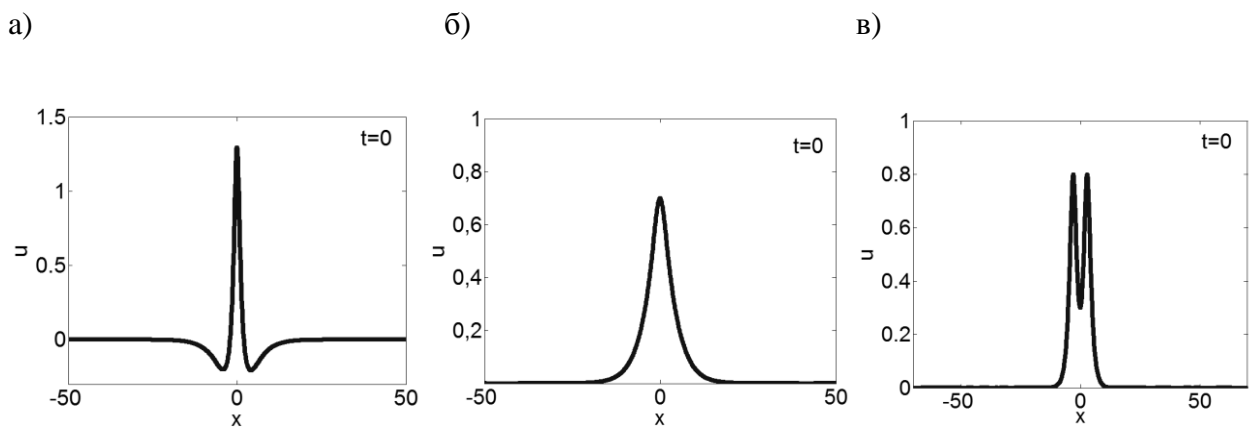

Рис. 1. Форма импульсов в момент взаимодействия солитонов: а) $s_1A_1=1$, $s_2A_2=-0.3$, б) $s_1A_1=1$, $s_2A_2=0.3$, в) $s_1A_1=1$, $s_2A_2=0.7$

Важно отметить, что амплитуда одногорбового результирующего импульса легко находится из (5) при $x = 0$:

$$U_* = A_1 - s_2 A_2, \qquad (8)$$

так что амплитуда растет при взаимодействии солитонов разной полярности и убывает при взаимодействии солитонов одинаковой полярности.

В случае двугорбового результирующего импульса, его максимальное значение находится не в центральной точке, и оно находится из (5) в весьма громоздком виде.

На рис. 2 приведены величины максимального и минимального значений результирующего импульса от отношения амплитуд солитонов для четырёх видов взаимодействия (три вышеупомянутых, а четвёртый - для знакопеременных солитонов, когда наибольший имеет отрицательную полярность). Несложно заметить симметрию рисунков 2а и 2г, а так же 2б и 2в. Это объясняется одинаковыми амплитудами солитонов, но противоположностью знаков обоих солитонов.

Таким образом, в случае 2а и 2г амплитуда результирующего импульса сначала монотонно убывает (возрастает) до значения $A_2/A_1 = 0.41,$ при этом принимая значение $|0.607|$ (в данном случае амплитуда второго солитона мала и результирующий импульс имеет одногорбовую форму, амплитуда которого находится из формулы (8)). Затем опять монотонно возрастает (убывает) (при больших амплитудах второго солитона, когда результирующий импульс двугорбый). Такое поведение объясняется сменой режимов обгонного на обменный для отношения $0.38 < A_2/A_1 < 0.43$. Фактически здесь полная аналогия с динамикой КдВ солитонов, где также есть переходная зона между двумя режимами [24].

В случае разнополярных солитонов (рис.2б, в) кривые изменения положительных и отрицательных амплитуд суммарного импульса монотонные, так как тут имеет место только один режим взаимодействия солитонов. Максимум импульса в момент взаимодействия для рис. 2б и, соответственно, минимум для рис. 2г, убывает (возрастает) с уменьшением модуля амплитуды второго солитона и равна она $s_1 A_1 - s_2 A_2$ (по аналогии с (8)).

а)   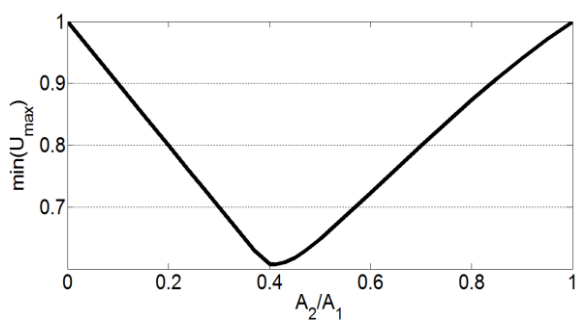   б) 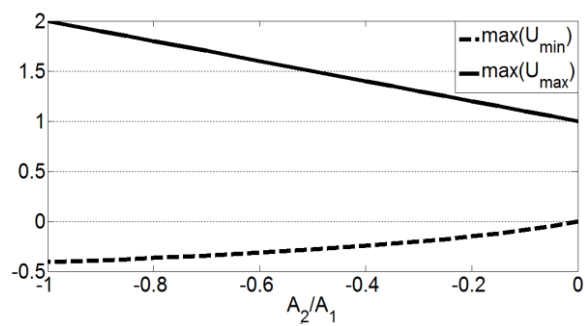

в) 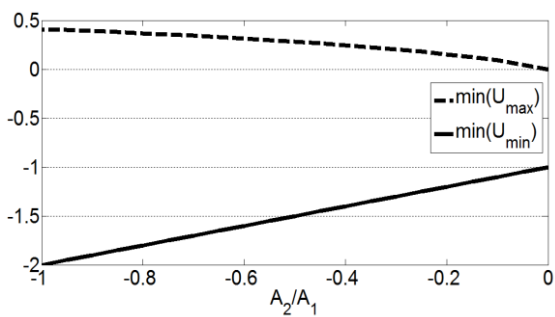   г) 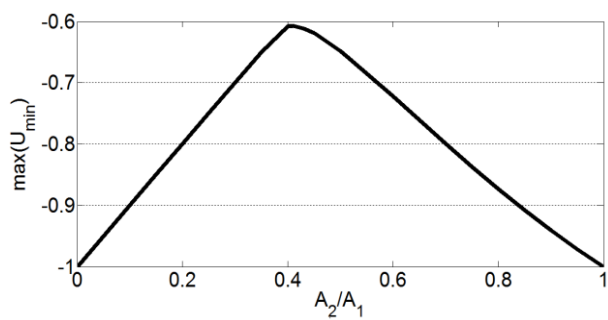

Рис. 2 Экстремумы волновых полей: a) положительные мКдВ-солитоны, b) знакопеременные мКдВ-солитоны (больший солитон имеет положительную полярность), c) разнополярные мКдВ-солитоны (больший солитон имеет отрицательную полярность), d) отрицательные мКдВ-солитоны

## 2 ИНТЕГРАЛЬНЫЕ ХАРАКТЕРИСТИКИ

Как известно, модифицированное уравнение Кортевега - де Вриза является полностью интегрируемым, и оно имеет бесконечное количество сохраняющихся инвариантов [10], первые три из них соответствуют законам сохранения массы, момента и энергии:

$$I_1 = \int_{-\infty}^{+\infty} u\,dx, \qquad I_2 = \int_{-\infty}^{+\infty} u^2\,dx, \qquad (9)$$

$$I_3 = \int_{-\infty}^{+\infty}\left[u^4 - u_x^2\right]dx, \qquad I_4 = \int_{-\infty}^{+\infty}\left[u^6 - 5u^2 u_x^2 + \frac{1}{2}u_{xx}^2\right]dx, \qquad (10)$$

Эти интегралы сохраняются в процессе эволюции волнового поля и их легко вычислить аналитически для случая, когда солитоны разделились в пространстве:

$$I_1 = \pi\left(1 + s_2(A_2)\right) \qquad I_2 = 2(A_1 + A_2) \qquad (11)$$

$$I_3 = \frac{2}{3}\left(A_1^3 + A_2^3\right), \qquad I_4 = \frac{A_1^5 + A_2^5}{5} \qquad (12)$$

Величина первого инварианта играет важное значение в эволюции начального возмущения, определяя число возникающих солитонов и бризеров, она может быть как положительна, так и равной нулю [26]. Остальные инварианты оказываются положительно определенными, и их величины растут с увеличением амплитуд взаимодействующих солитонов независимо от их полярности. Знание этих интегралов важно в первую очередь для контроля численных решений модифицированного уравнения Кортевега - де Вриза.

Так как в аналитике показано, что основной вклад в динамику мультисолитонных вносят парные взаимодействия солитонов [19, 27], для понимания этого вклада и влияния подобных взаимодействий на статистические моменты полнового поля, мы исследовали следующие интегралы, соответствующие четырём статистическим моментам:

$$M_n(t) = \int_{-\infty}^{+\infty} u^n(x,t)\,dx, \; n = 1,2,3,4 \qquad (13)$$

Первые два интеграла будут сохраняться с течением времени в силу интегрируемости уравнения мКдВ. Однако третий и четвёртый моменты, соответствующие асимметрии и эксцессу в теории турбулентности, не являются инвариантами и изменяются во времени (рис. 3). В случае взаимодействия двух положительных солитонов третий и четвёртый моменты убывают, как и в аналогичной задаче для КдВ-солитонов [23, 24] (рис. 3а). Физически это можно объяснить эффектом уменьшения амплитуды результирующего импульса в момент взаимодействия. Для взаимодействия же солитонов разных полярностей, описываемых формулой (3), когда больший солитон остаётся положительным, а меньший становится отрицательным – наоборот, амплитуда суммарного импульса значительно вырастает, и это даёт вклад в изменения третьего и четвёртого моментов и они оба увеличиваются в момент взаимодействия (рис. 3б). В случае взаимодействия двух отрицательных солитонов третий момент отрицателен и увеличивается в момент взаимодействия, а четвёртый уменьшается (рис. 3г). В случае же, если больший солитон, находящийся слева, имеет отрицательную амплитуду, а меньший солитон – положительную, третий момент по-прежнему отрицателен и уменьшается в момент взаимодействия, а четвёртый увеличивается (рис. 3в).

a) 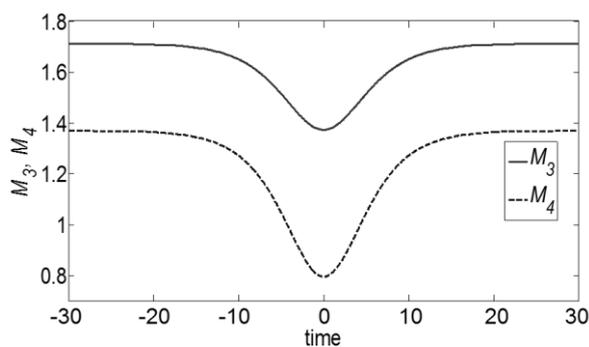

b) 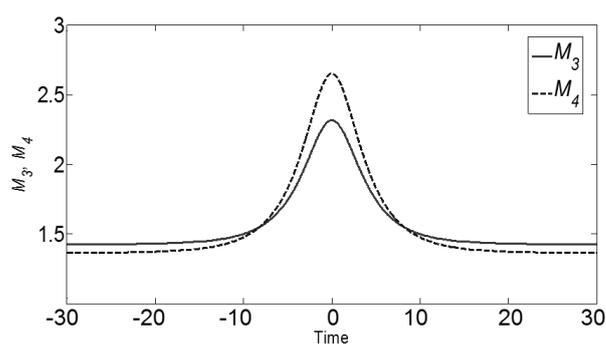

c)

d)

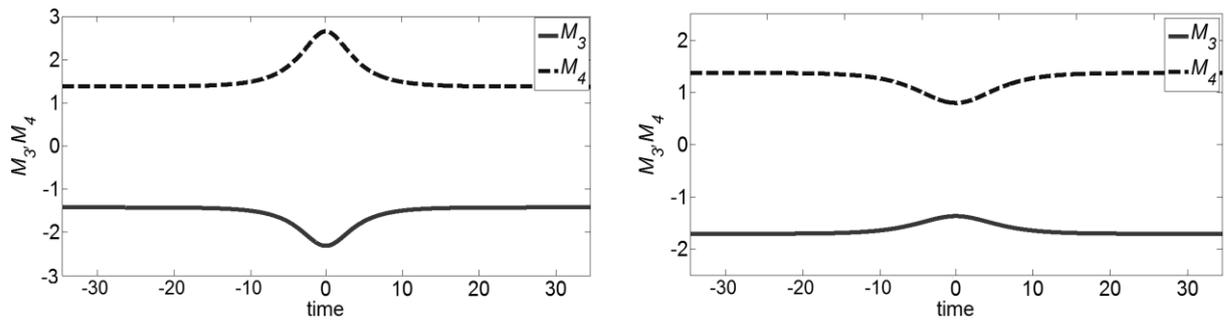

Рис. 3. Зависимость моментов $M_3$ и $M_4$ от времени при взаимодействии мКдВ-солитонов: а) $A_1$=1, $A_2$=0.3, б) $A_1$=1, $A_2$=-0.3, в) $A_1$=-1, $A_2$=0.3, г) $A_1$=-1, $A_2$=-0.3.

Для невзаимодействующих солитонов эти интегралы можно вычислить аналитически:

$$M_1 = \pi(s_1 A_1 + s_2 A_2), \qquad M_2 = 2(A_1 + A_2),$$

$$M_3 = \frac{\pi}{2}(s_1 A_1^2 + s_2 A_2^2), \qquad M_4 = \frac{4}{3}(A_1^3 + A_2^3) \qquad (14)$$

Формулы (14) определяют начальные и конечные значения моментов, когда солитоны разделены. Таким образом, знак первого и третьего моментов зависит от полярностей солитонов. Второй момент, как и четвертый, всегда положителен.

Чтобы оценить величину изменений моментов при взаимодействии солитонов, рассмотрим изменение величин третьего и четвертого моментов $M_3^*, M_4^*$ в зависимости от отношения амплитуд солитонов. Здесь $M_i^* = (M_{i\_\max} - M_{i\_\min})/M_{i\_0}$. Опять присутствует симметрия для третьих моментов на рис. 4а и 4г и для 4б и 4в, а четвёртые моменты одинаковы для соответствующих графиков.

Для однополярных солитонов (рис. 4 а,г), как и ранее, наблюдается смена режимов взаимодействия солитонов. Величина изменения моментов максимальна как раз для солитонов с отношением $A_2/A_1$ соответствующим переходной зоне, и изменения моментов в этом случае может достигать 20% и 40%, соответственно, для третьего и четвёртого моментов.

Так же на рис. 4а для сравнения представлены соответствующие кривые для КдВ-солитонов, которые лежат немного ниже кривых для мКдВ-солитонов.

В случае взаимодействия разнополярных солитонов характер кривых значительно меняется (рис. 4б, в). Кривые монотонны, так как в данном случае существует только один режим взаимодействия. Важно отметить, что величина изменений моментов для разнополярных солитонов весьма существенна, особенно когда амплитуды солитонов близки по модулю.

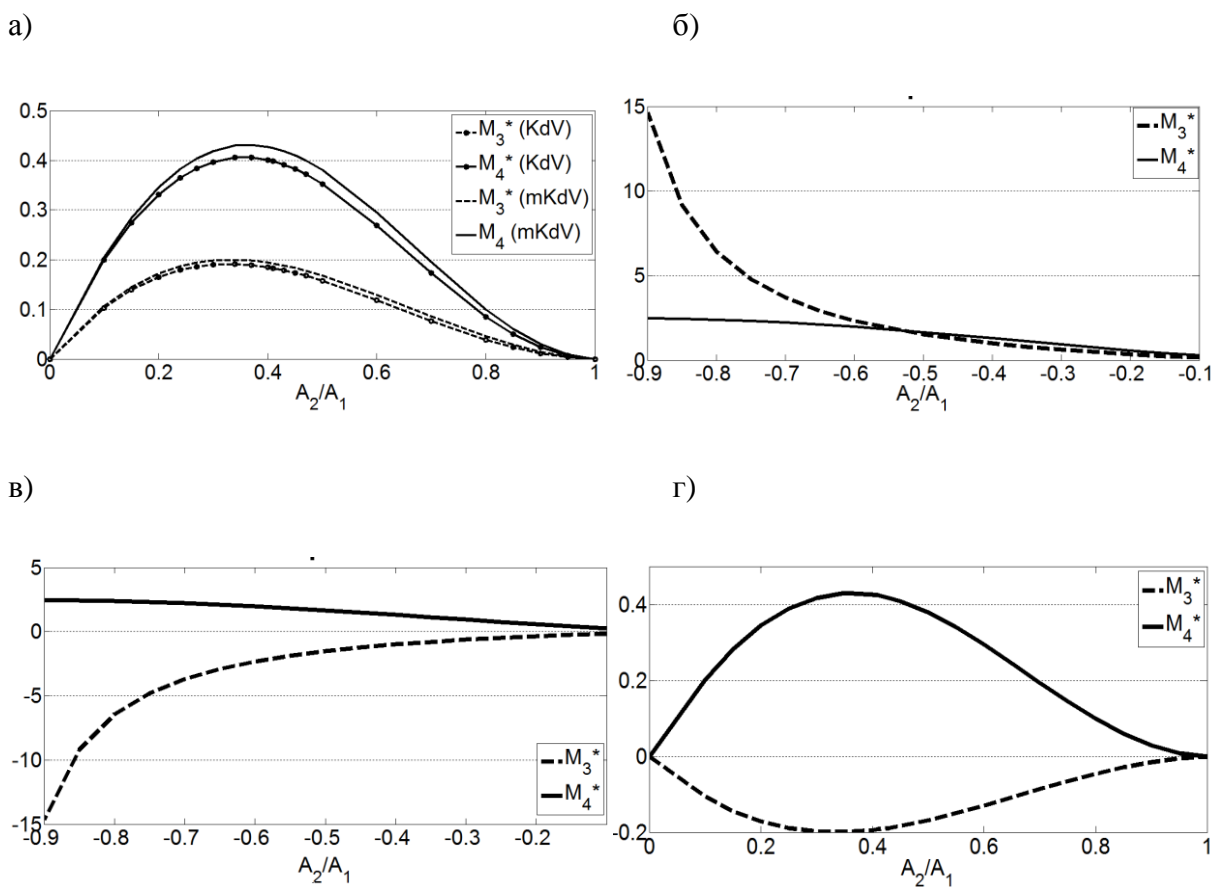

Рис. 4. Изменение величин третьего и четвертого моментов $M_3^*, M_4^*$ в зависимости от отношения амплитуд солитонов: а) положительные мКдВ-солитоны и КдВ-солитоны с соответствующими амплитудами, б) знакопеременные мКдВ-солитоны (больший солитон имеет положительную полярность), в) разнополярные мКдВ-солитоны (больший солитон имеет отрицательную полярность), г) отрицательные мКдВ-солитоны.

Таким образом, взаимодействие двух солитонов сильно влияет на моменты волнового поля, так что этот эффект может оказаться важным для понимания природы солитонной турбулентности.

**3 ЗАКЛЮЧЕНИЕ**

В данной работе изучена динамика двухсолитонного поля в рамках модифицированного уравнения Кортевега де - Вриза. Детально исследован процесс соударения двух солитонов. Обсуждены возможные режима взаимодействия солитонов. Вычислены первые четыре момента волнового поля, играющие важную роль в теории турбулентности. Первые два из них являются интегралами движениями для модифицированного уравнения Кортевега – де Вриза, и они сохраняются. Показывается, что взаимодействие солитонов одинаковой полярности ведёт к уменьшению третьего и четвертого моментов, характеризующих асимметрию и эксцесс волнового процесса (как и в случае КдВ уравнения), однако взаимодействие солитонов разной полярности ведёт к увеличению этих моментов солитонного поля. В случае взаимодействия положительных солитонов наибольший вклад в динамику моментов вносят солитоны с отношением амплитуд $0.34 < A_2/A_1 < 0.43$ (переходный режим между обменным и обгонным взаимодействием), а для знакопеременных солитонов - солитоны с отношением амплитуд, стремящимся по модулю к единице. Отсюда становится ясным, что двухсолитонные взаимодействия могут сильно влиять на моменты волнового поля, и это является важным фактором для понимания солитонной турбулентности.

**Список литературы**

**Подписи к рисункам**

а)                                    б)                                    в)

Рис. 1. Форма импульсов в момент взаимодействия солитонов: а) $s_1A_1=1$, $s_2A_2=-0.3$, б) $s_1A_1=1$, $s_2A_2=0.3$, в) $s_1A_1=1$, $s_2A_2=0.7$

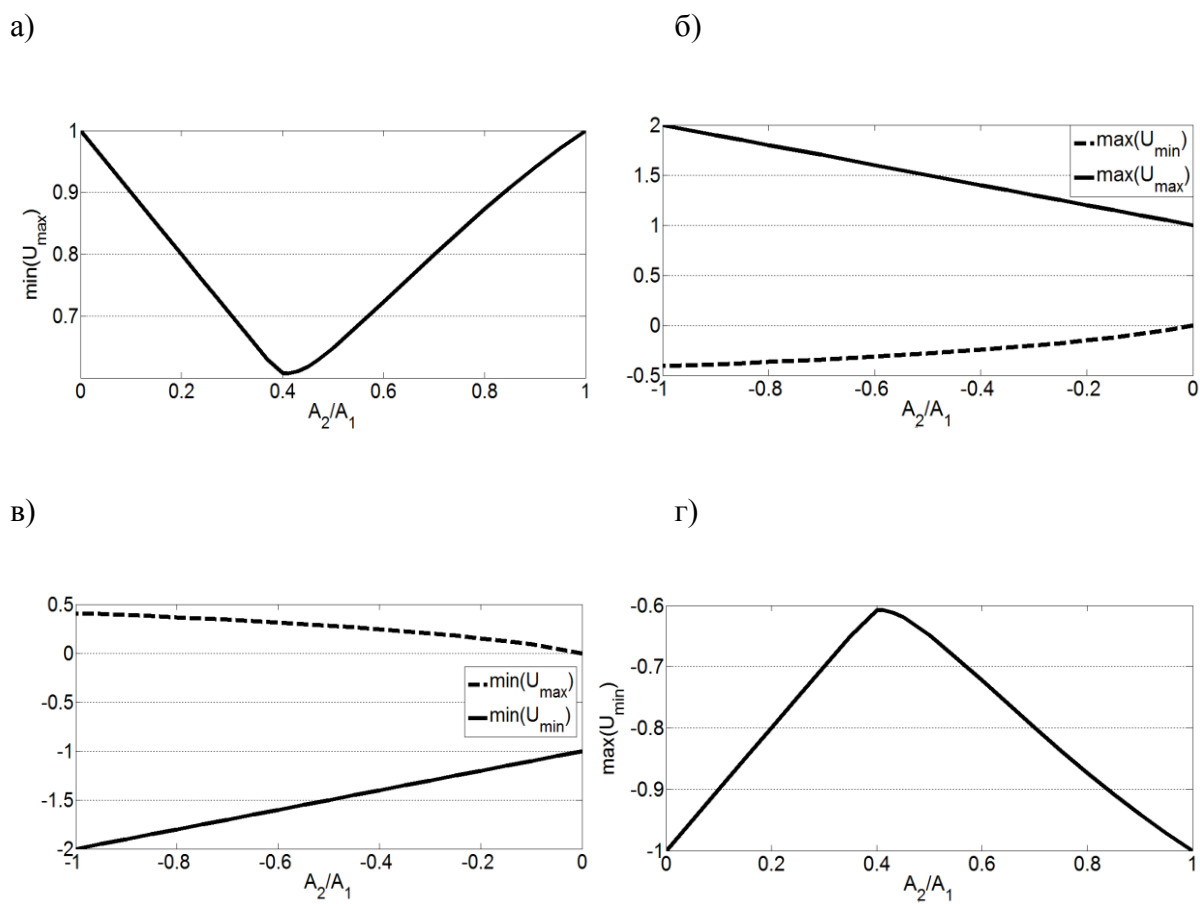

Рис. 2 Экстремумы волновых полей: a) положительные мКдВ-солитоны, b) знакопеременные мКдВ-солитоны (больший солитон имеет положительную полярность), c) разнополярные мКдВ-солитоны (больший солитон имеет отрицательную полярность), d) отрицательные мКдВ-солитоны

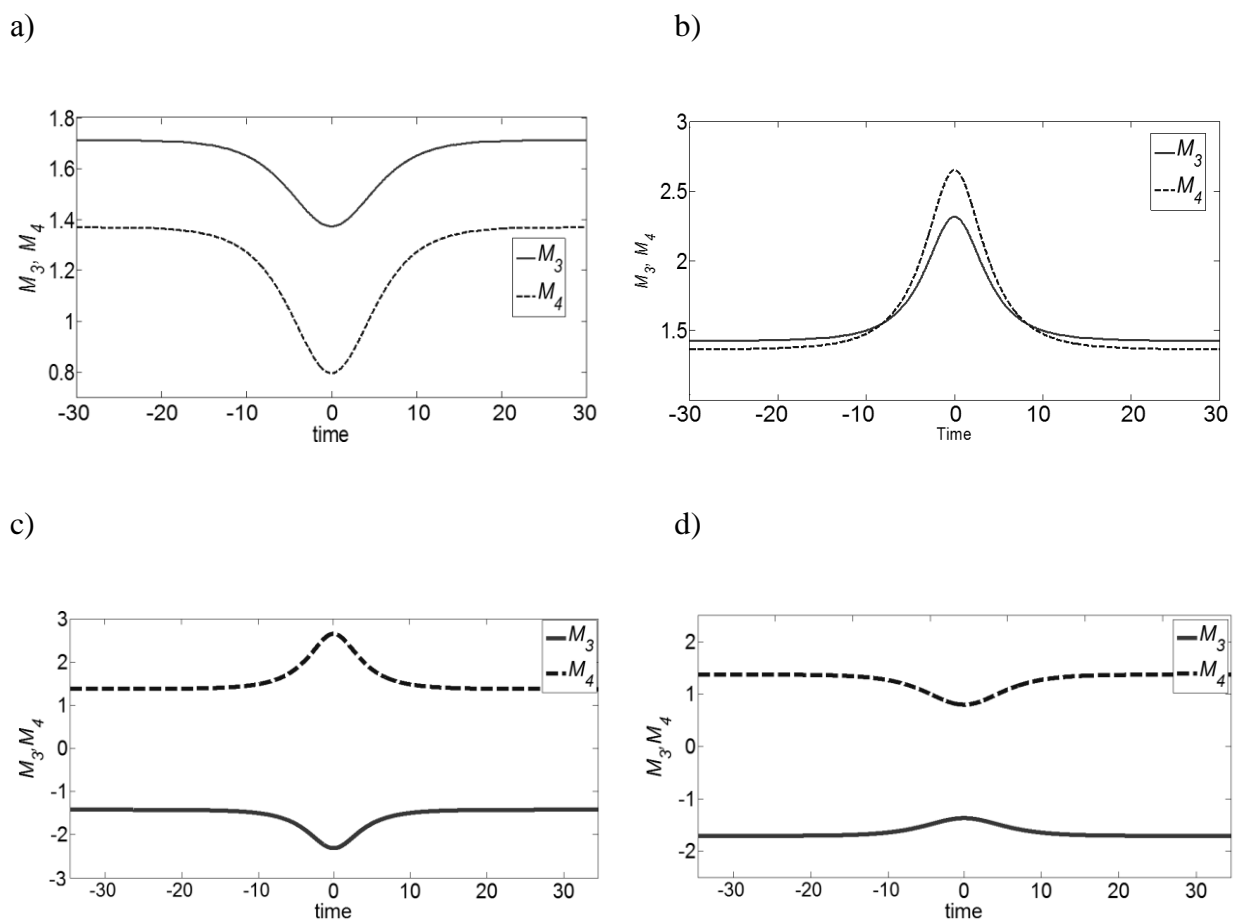

Рис. 3. Зависимость моментов $M_3$ и $M_4$ от времени при взаимодействии мКдВ-солитонов: а) $A_1$=1, $A_2$=0.3, б) $A_1$=1, $A_2$=-0.3, в) $A_1$=-1, $A_2$=0.3, г) $A_1$=-1, $A_2$=-0.3.

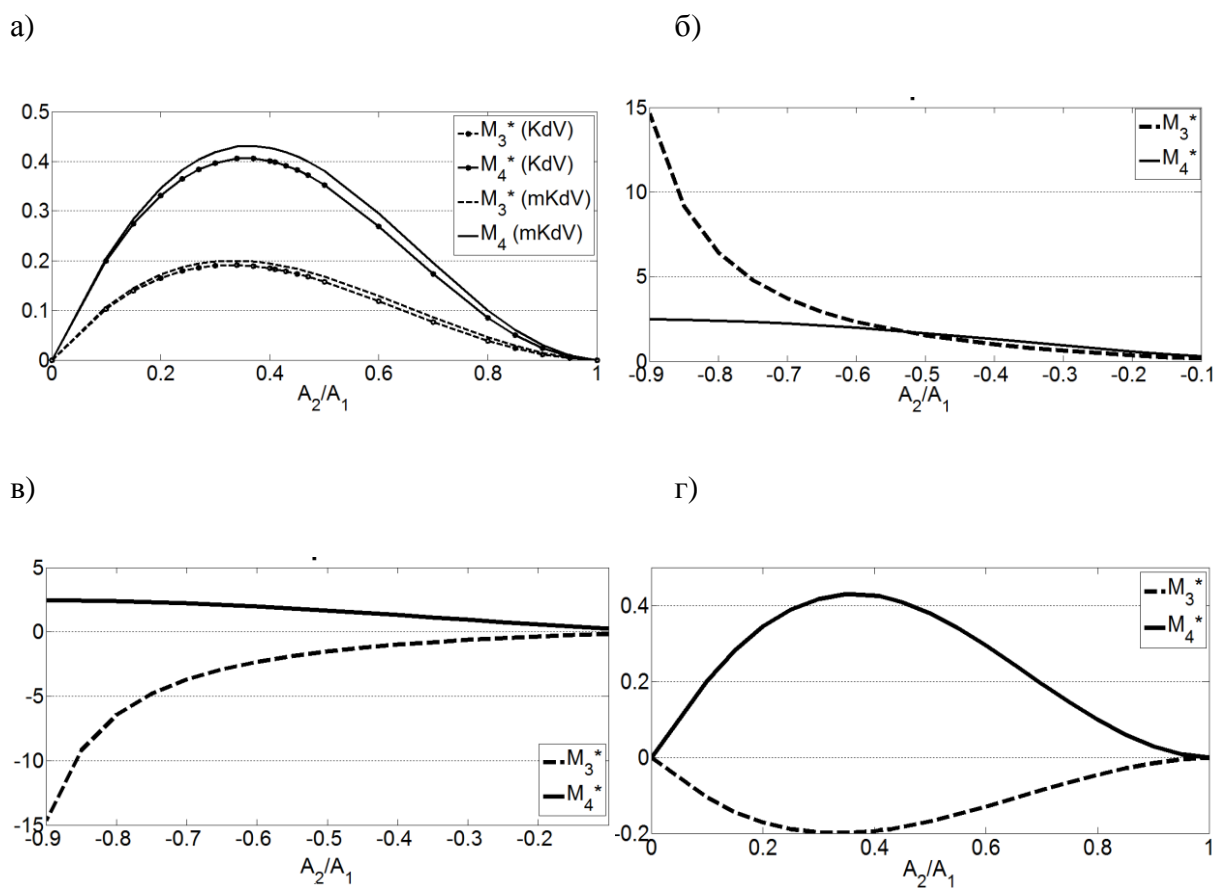

Рис. 4. Изменение величин третьего и четвертого моментов $M_3^*, M_4^*$ в зависимости от отношения амплитуд солитонов: а) положительные мКдВ-солитоны и КдВ-солитоны с соответствующими амплитудами, б) знакопеременные мКдВ-солитоны (больший солитон имеет положительную полярность), в) разнополярные мКдВ-солитоны (больший солитон имеет отрицательную полярность), г) отрицательные мКдВ-солитоны.